\documentclass[referee]{aa}
\usepackage{epsfig}
\usepackage{graphicx}
  
\begin{document}
\title{Gravitational wave background from magnetars}
\author{Tania Regimbau \and Jos\'e Antonio de Freitas Pacheco}
\offprints{T. Regimbau, \email{regimbau@obs-nice.fr}}
\institute{UMR 6162 Artemis, CNRS, Observatoire de la C\^ote d'Azur, BP
  4229, 06304 Nice Cedex 4, France \and UMR 6202 Cassiop\'ee, CNRS, Observatoire de la C\^ote d'Azur, BP 4229, 06304 NICE Cedex 4 (France)}
\date{Received \today / Accepted} 

\abstract{
We investigate the gravitational wave background produced by
magnetars. The statistical properties of these highly magnetized
stars were derived by population synthesis methods and assumed to be
also representative of extragalactic objects. The adopted ellipticity
was calculated from relativistic models using equations of state and assumptions concerning the distribution of
currents in the neutron star interior. The maximum amplitude occurs
around 1.2 kHz, corresponding to $\Omega_{gw} \sim 10^{-9}$ for a type
I superconducting neutron star model. The expected signal is a continuous background 
that could mask the cosmological contribution produced in the early stage of the Universe.

}
\titlerunning{GW from magnetars}

\maketitle

\section{Introduction}

The stochastic background is a rich source of gravitational
waves (GW). The detection of a cosmological gravitational background (CGB) is an
unique way to probe the very early stages of the universe up to the limits of the Plank
era and the big bang (Grishchuk et al., 2001). In addition to the CGB, the emission
from a large number of unresolved sources is
expected that produces the astrophysical gravitational background (AGB).
Because the AGB mask the relic signal, the knowledge of its properties
such as
the spectral energy distribution is of fundamental importance to determine the best
frequency domain in which to search for the CGB in order to optimize the subtraction techniques.
A large number of astrophysical processes able to generate
a stochastic background have been investigated. On the one hand,
distorted black holes (Ferrari et al. 1999a; de Ara\'ujo et al. 2000)
and  bar mode 
emission from young neutron stars
(Regimbau 2001) are examples of sources able to generate a shot noise signal (the time interval
between events is large in comparison to the duration of a single event), while supernovas
(Blair et al. 1997; Coward et al. 2000; Buonanno et al. 2004) are expected to produce an
intermediate ``popcorn'' noise. On the other hand, the contribution of tri-axial rotating
neutron stars results in a truly continuous background that may compete with the relic
one in the frequency domain of ground based interferometers (Regimbau \& de Freitas Pacheco
2001b, hereafter RP01b).
Neutron stars (NS) are one of the most promising sources for
observation by ground based
gravitational wave interferometers such as LIGO, VIRGO, GEO and TAMA
(see de Freitas Pacheco 2001 for a recent review). 
Various processes related to NS able to produce large numbers
of gravitational waves (GWs) have been investigated: free precession of an axisymmetric star induced by a
misalignment between the spin and the symmetry axes (Zimmermann \&
Szedenits 1979), r- or bar-mode instabilities excited by the so-called
CFS (Chandrasekhar-Friedman-Schultz) mechanism, which may occur in
fast and hot rotating neutron stars (Andersson \& Kokkotas 2001 and
references therein) and mini gravitational collapse induced by a phase
transition in the core (Marranghello, Vasconcellos \& de Freitas
Pacheco 2002). However, the emission produced by a tri-axial rotating
star is still one of the most studied mechanisms.
Different scenarios leading to a distorted star have been discussed in the
literature, as anisotropic stresses from strong magnetic fields and
tilting of the symmetry axis during the initial cooling phase when the
crust solidifies. Bildsten (1998) pointed out that a neutron star in a
state of accretion may develop non-axisymmetric temperature variations
on the surface, which produce horizontal density patterns able to
create a large mass quadrupole moment, if the elastic response of
the crust is neglected. More detailed  calculations (Ushomirsky,
Cutlerand Bildsten 2000) indicate that the  inclusion of the crustal
elasticity decreases the expected mass quadrupole by a factor of 20-50, reducing 
considerably the predicted GW emission. More  recently, elastic deformations
of compact objects with ``exotic" equations of state have been considered
by Owen (2005). He showed that solid strange stars could sustain ellipticities
as high as few times $10^{-4}$, considerably higher than estimated values for
conventional neutron stars (Thorne 1980).
The distortion induced by magnetic fields becomes significant in
highly magnetized neutron stars, overwhelming the ``flattening"  due
to a fast rotation. The existence of neutron stars with magnetic fields
in excess of $10^{14} G$ (called magnetars) was firstly suggested  
by Thompson \& Duncan (1992). They have shown that in fast newborn neutron stars, the dynamo 
mechanism could generate 
magnetic field strengths up to $10^{16}$ G. The existence of magnetars
is supported by the observation of ``soft gamma repeaters" (SGRs) and
anomalous X-ray pulsars (AXPs), whose rotation periods and
deceleration rates, if interpreted in terms of the canonical magnetic 
dipole model, suggest that these objects are associated
with young highly magnetized NS (Kouveliotou et al. 1998, 1999; de Freitas Pacheco 1998;
 Heyl \& Kulkarny 1998; Mereghetti 1999).
On the one hand, magnetars may be considered as objects having an origin and
evolutionary path different from ``classical" radio-pulsars but, on the
other hand, simulations based on population synthesis methods suggest
that NS are born with a large variety of rotational periods,
magnetic fields and that magnetars are objects simply born in the high
end of the magnetic field distribution (Regimbau \& de Freitas
Pacheco 2001a, hereafter RP01a). According to these simulations, a non-negligible fraction of NS are born with magnetic fields higher than $10^{14}$ G, raising 
the possibility that these objects may also generate a continuous
stochastic background.
In RP01b, the contribution of
tri-axial rotating neutron stars to the gravitational wave background
was calculated under the assumption that the initial period and magnetic field
distributions derived from population synthesis of galactic pulsars by
Regimbau \& de Freitas Pacheco (2000, hereafter RP00) could be extended to the
extragalactic population. In this paper, we adopt a similar procedure
to compute  the contribution of magnetars to the GW background, but
with a fundamental difference. In RP00 the equatorial ellipticity was
taken as a free parameter, whereas in the present work the equatorial
ellipticity is estimated from NS models including different magnetic
field configurations and equations of state. This paper is organized
as follows: in Section 2 we discuss the properties of the magnetar population, including
the magnetic field distribution and the equatorial deformation, in
Section 3 their contribution to the gravitational wave background is calculated, in Section 4
the expected signal is compared to the CGBs and the detection possibility by present and future generations 
of GW detectors is discussed and in Section 5 the main conclusions are summarized.

\section{Properties of magnetars}

\subsection{The magnetic field distribution}

In our previous work (RP00), the population synthesis method was used to recover the statistical properties
of the real population of radio pulsars, since most of the objects are hidden by strong
selection effects, such as the flux density limit of radio telescopes and their
orientation with respect to the pulsar emission beam. 
These simulations were recently upgraded in order to take into account
the new pulsars discovered by the 
Parkes multibeam survey. In agreement with our previous analysis, the initial period follows a 
normal distribution of mean $240$ ms with a dispersion of $80$ ms, while the magnetic field has 
log-normal distribution of mean log $B$ = 13.0 (in Gauss) with a dispersion of $\sigma_{log B} =$0.8. For 
details of these simulations, the reader is referred to RP00.

%
%
\begin{figure}
\resizebox{\hsize}{!}{\includegraphics[width=5in]{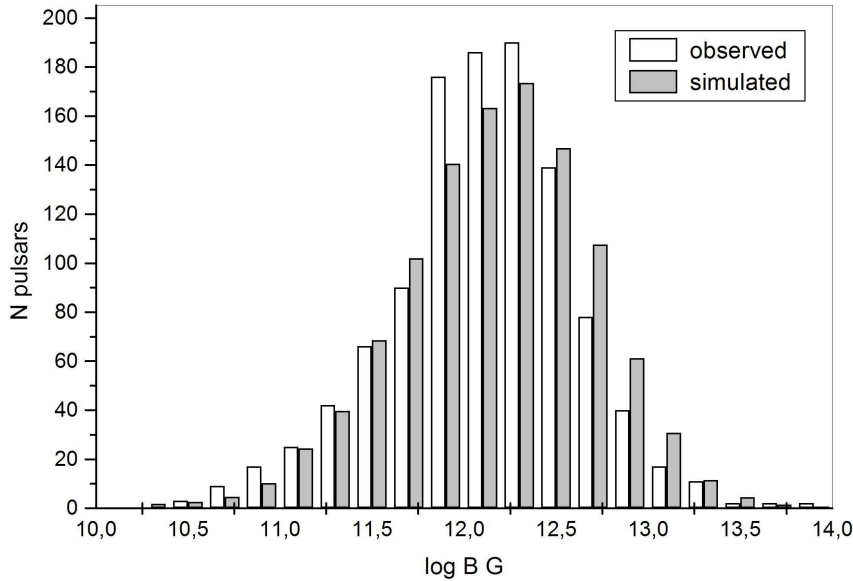}}\hfill
\caption{Comparison between the simulated and the observed distribution (derived from
the ``canonical" magnetic dipole model) of the magnetic field (in Gauss).}
\end{figure}

As shown in Fig. 1, the magnetic field distribution of the
observed population matches quite well that derived from the observed period deceleration rate
and the ``canonical" magnetic dipole model. In this case, the mean field of the observed
population is about $1.6 \times 10^{12}$ G and the distribution has an apparent
cutoff around $B = 10^{14}$ G. As we have already stated (RP01a),
the average magnetic field of the real (or ``hidden") population is one
order of magnitude higher and the fraction of magnetars, defined as
objects with fields above $10^{14}$ G, is around $8\%$. A comparison between the magnetic
field distributions of the observed and the ``unseen" populations  is given in Fig. 2. Notice
that objects with fields as high as $3 \times 10^{15}$ are expected to exist in the Galaxy.

%
%
\begin{figure}
\resizebox{\hsize}{!}{\includegraphics[width=5in]{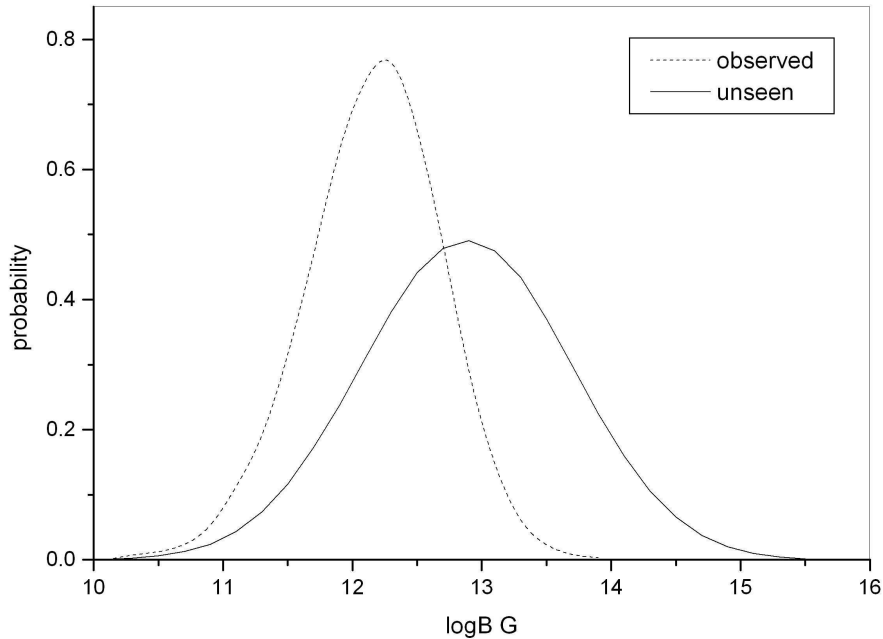}}\hfill
\caption{Magnetic field distributions of the observed
  (dashed line) and the unseen population (full line) of pulsars}
\end{figure}

\subsection{Deformation by magnetic fields}

The deformation of magnetized Newtonian stars was already discussed in
the early fifties by Chandrasekhar \& Fermi (1953) and Ferraro
(1954). The gravitational wave emission from magnetic distorted stars
was considered by Gal'tsov \& Tsvetkov (1984), Bonazzola \& Gourgoulhon (1996), Konno, Obata \& Kojima 
(1999, 2000), Palomba (2001), among others.
The deformation of a slowly rotating magnetized star can be expressed
as the sum of the contribution of three main terms: the first and the
most important corresponds to the Lorentz force, induced by current
flows in the highly conductive NS interior; the second term represents
variations of the gravitational potential, a consequence of the
distortion itself and the third is a purely relativistic term arising
from the definition of the circumferential radius $R = L/(2\pi)$, where $L$ is
the length of the equator as measured by a non-rotating observer. In general, the ellipticity induced by 
magnetic field effects can be expressed by the dimensionless ratio (Konno, Obata \& Kojima 2000)

\begin{equation}
\varepsilon_B = g\frac{B^2R^4}{GM^2}\mathrm{sin}^2\alpha 
= 1.9\times 10^{-8}gB^2_{14}R^4_{10}M^{-2}_{1.4}\mathrm{sin}^2\alpha
\end{equation}
where B, R and M are respectively the magnetic field at the surface,
the radius and the mass of the star, $\alpha$ is the angle between the
spin and magnetic dipole axes, while $g$ is a dimensionless parameter depending on both 
the equation of state (EOS) and on the magnetic field geometry. It is equivalent to the parameter $\beta$ introduced by Bonazzola \& Gourgoulhon (1996).

In the case of an incompressible fluid star with a dipole magnetic
field, considered by Ferraro (1954), the deformation parameter is
$g$ = 12.5. Relativistic models based on a polytropic EOS, e.g., $P
\propto \rho^{\Gamma}$ and a dipole field geometry give similar values
(Konno, Obata \& Kojima 2000). NS models built with the EOS
UV$_{14}+$TNI (Wiringa, Fiks \& Fabrocini 1988) and with non-superconducting neutron star matter lead to deformation parameters $g$
of the order of the unity for currents concentrated in the crust and
a few times higher if currents are concentrated in the 
core (Bonazzola \& Gourgoulhon 1996). These values correspond to ellipticities of 
about $(2- 20)\times 10^{-8}$ for fields typically of the order of $10^{14}$ G, below the limits 
estimated by RP01b to produce a detectable signal by the present generation of interferometric detectors.
A more favourable situation occurs if neutron stars have a superconducting interior of type I. In 
this case, the magnetic field permeates only the very outer layers of the star and distortion 
parameters as high as $g \approx$ 520 can be obtained (Bonazzola \& Gourgoulhon 1996). 
From eq. 1 and the magnetic field distribution of the actual
population derived in the previous section, the ellipticity
distribution of magnetars can be calculated (Fig.3).

%
%
\begin{figure}
\resizebox{\hsize}{!}{\includegraphics[width=5in]{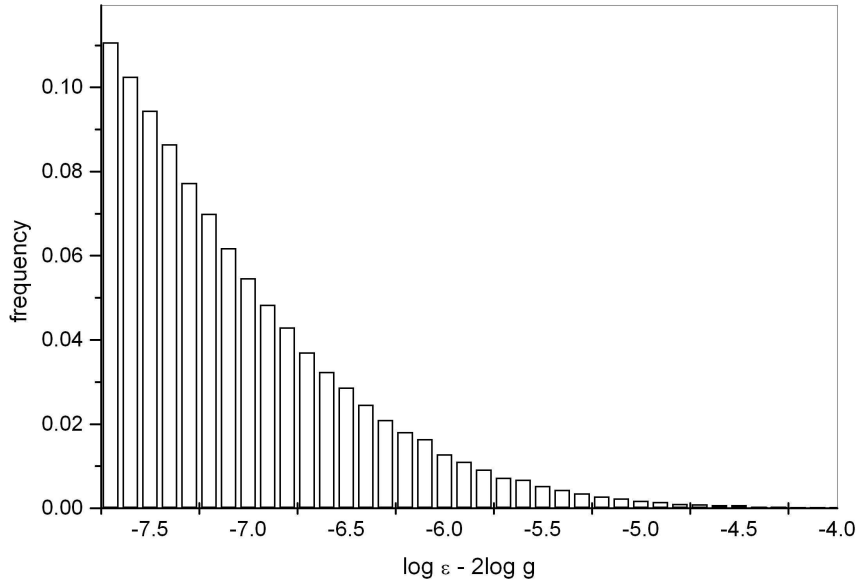}}\hfill
\caption{Simulated distribution probability for the expected magnetic ellipticity of magnetars}
\end{figure}

Our simulated data can be quite well fitted by the function
\begin{equation}
{\rm p(x)} = 2.36 \,\ {\rm exp}(-0.2({\rm x} + 9.7)^2)\,\ {\rm if}\,\ {\rm x} > -7.7
\end{equation}
where $\mathrm{x} = \mathrm{log}(\varepsilon_B) - 2 \mathrm{log}(g)$

According to our simulations, even if magnetars with ellipticities as high 
as $\varepsilon_B \sim 0.01$ are expected to exist, their gravitational radiation remains 
small, since most of them have low rotation frequencies (RP01a). This implies that in spite of
their high ellipticity, these objects are always decelerated by the magnetic braking mechanism, a
point to be retained when the spectral energy is computed (see
Section 3). To illustrate 
this point, the distribution of the ratio $R$ between the two energy loss mechanisms, namely
\begin{equation}
R=\frac{\dot{E}_{gw}}{\dot{E}_{md}}=1.36\times 10^{-12}(\frac{g}{13})^2 B_{14}^2\mathrm{sin}^2\alpha P^{-2}
\end{equation} 
is shown in Fig. 4. The numerical value was calculated for a ``canonical" NS of mass M = 1.4 M$_{\odot}$,
radius R = 10 km, moment of inertia I = 1.4$\times 10^{45}$ gcm$^2$ and B = 10$^{14}$ G. 
Notice that even for ``extreme" cases, e.g., magnetars in the tail of the
period and magnetic field distributions, the aforementioned ratio is $R \sim 0.002 << 1$
for objects with $P \sim$ 1ms, $B_{14} \sim$ 1 and $g \sim 520$. 

%
%
\begin{figure}
\resizebox{\hsize}{!}{\includegraphics[width=5in]{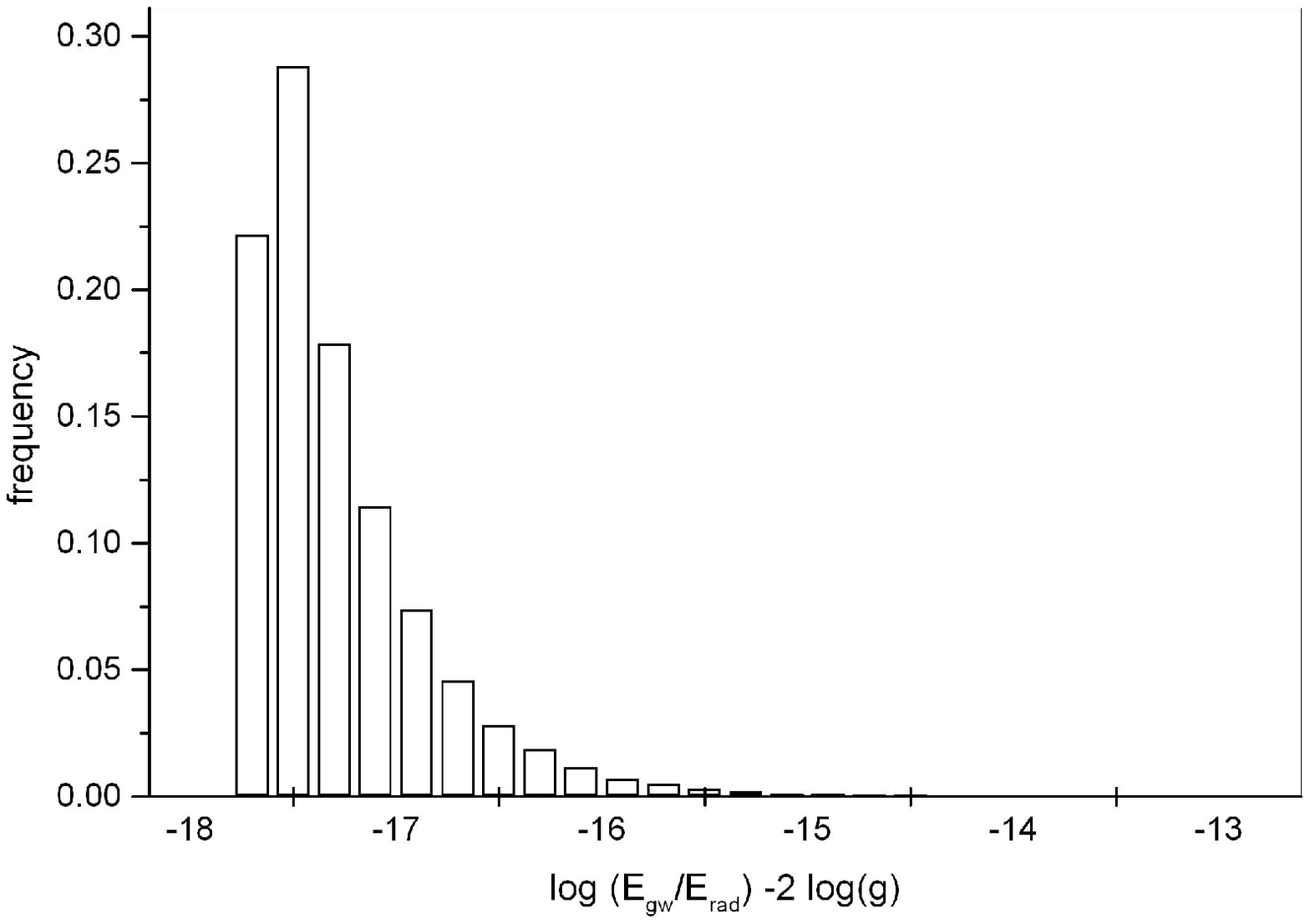}}\hfill
\caption{The ratio between the gravitational and magnetic energy loss rates for magnetars}
\end{figure}

\section{The stochastic gravitational wave background}

\subsection{Spectral properties}

The superposition of the gravitational wave emission from different
objects and in
particular from magnetars formed since the beginning of the star
formation in the universe produces a stochastic background whose detection might be possible by
employing cross correlation techniques between two antennas (Grishchuk 1976; Christensen, 1997).

The spectral properties of the stochastic background are characterized
by the dimensionless parameter (Ferrari et al. 1999):
\begin{equation}
\Omega _{gw}(\nu _{o})=\frac{1}{c^{3} \rho _{c}}{\nu _{o}}F_{\nu _{o}}
\end{equation}
where $\nu _{o}$ is the wave frequency in the observer frame, $\rho _{c}$ is the critical mass 
density needed to close the Universe, related to the Hubble parameter $H_{0}$ by
\begin{equation}
\rho _{c}=\frac{3H_{o}^{2}}{8\pi G}.
\end{equation}
$F_{\nu_o}$ is the gravitational wave flux at the observer frequency $\nu_o$ integrated over 
all sources defined by
\begin{equation}
F_{\nu _{o}}=\int_{0}^{z_{\max }}f_{\nu _{o}}dR(z)
\end{equation}
where $f_{\nu_o}$ is the gravitational flux of a single source located
between $z, z+dz$ and $dR(z)$ is the source formation rate. Both
quantities are given by
\begin{equation}
f_{\nu _{o}}=\frac{1}{4\pi d_{L}^{2}}\frac{dE_{gw}}{d\nu }(1+z)
\end{equation}

and

\begin{equation}
dR(z)=\lambda _{p}\frac{R_{c}(z)}{1+z}{\frac{{dV}}{{dz}}}dz
\end{equation}

In the above equations, $d_{L}=(1+z)r$ is the distance luminosity, $r$
is the proper distance, which depends on the adopted cosmology,
${dE_{gw}}/{d\nu}$ is the gravitational spectral energy emitted by a single source and 
$\nu=(1+z)\nu _{o}$ the frequency in the source frame. 
$R_{c}(z)$ is the cosmic star formation rate and  $\lambda_{p}$ is the
mass fraction converted into neutron star progenitors with magnetic fields higher than
$10^{14}$ G. The (1+z) term
in the denominator of eq. 8  corrects the star formation rate by the time dilatation 
due to the cosmic expansion. The element of comoving volume is given by
\begin{equation}
dV = 4\pi r^2{{c}\over{H_o}}{{dz}\over{E(\Omega_i,z)}}
\end{equation}
with
\begin{equation}
E(\Omega_i,z) =[\Omega_m(1+z)^3+\Omega_v]^{1/2}
\end{equation}
where $\Omega_m$ and $\Omega_v$ are respectively the present values of
the density parameters due to matter (baryonic and non-baryonic) and
the vacuum, corresponding to a non-zero cosmological constant. A ``flat" cosmological
model ($\Omega_m + \Omega_v = 1$) was assumed. Porciani \& Madau (2001) provide three models 
for the cosmic star formation rate (SFR) history up to redshifts
$z\sim 5$. Differences between these models are mainly due to the various corrections applied, in particular those due to 
extinction by the cosmic dust. In our computations, the three possibilities were considered. Since
the final results are not significantly different according to the adopted SFR history, we
present here only results derived from the second model, labelled SFR2.

In our calculation, we have taken $\Omega_m$ = 0.30 and $\Omega_v$ = 0.70,
corresponding to the so-called concordant model derived from
observations of distant type Ia supernovae (Perlmutter et al. 1999;
Schmidt et al. 1998) and the power spectra of the cosmic microwave
background fluctuations (Spergel et al. 2003). The Hubble parameter H$_o$ was taken to be 
$65\,\mathrm{km\,s}^{-1}\mathrm{Mpc}^{-1}$.

Combining the equations above one obtains:

\begin{equation}
\Omega _{gw}(\nu _{o})=\frac{8\pi G}{3c^{2}H_{o}^{3}}\lambda_{p}\, 
\nu_{o}\int_{0}^{z_{\sup }}\frac{dE_{gw}}{d\nu }\frac{R_{SFR2}(z)}{(1+z)^{7/2}}dz
\end{equation}
or numerically:
\begin{equation}
\Omega _{gw}(\nu _{o})=7.5 \times 10^{-56}h^{-2}_{65}\lambda_{p}\, 
\nu_{o}\int_{0}^{z_{\sup }}\frac{dE_{gw}}{d\nu }\frac{R_{SFR2}(z)}{(1+z)^{7/2}}dz
\end{equation}
 
In the above equation, the SFR (in $\mathrm{M}_{\odot}\mathrm{Mpc}^{-3}\mathrm{yr}^{-1}$) as
given by Porciani \& Madau (2001) was derived for a matter dominated
cosmology ($\Omega_m$ = 1) with H$_o = 65\,\mathrm{km\,s}^{-1}\mathrm{Mpc}^{-1}$, but 
eqs. (11) and (12) already include the correction for the adopted cosmology.
The upper limit of the integral is determined by the frequency cutoff in the
source frame, namely
\begin{equation}
z_{\sup }=\frac{\nu _{\sup }}{\nu_{o}} - 1
\end{equation}
Here we have assumed that the minimum rotation period is $\sim$ 0.8 ms, which corresponds to
a maximum gravitational wave frequency $\nu_{sup} \approx$ 2500 Hz.
The cosmic SFR is quite uncertain for $z > 5$, but if one imposes the
restriction $z_{sup} \sim 5$, there are no practical consequences since objects formed at higher redshifts
do not contribute significantly to the integral in eq. (12).

\subsection{Magnetar parameters}

The mass fraction of formed stars giving origin to magnetars is 
$\lambda_p = \chi\int^{40}_{10}\xi(m)dm$, where $\xi(m) \propto
m^{-2.35}$ is the initial mass function, supposed to be given by
Salpeter's law, and NS progenitors are supposed to form in the mass range 10-40
M$_{\odot}$. $\chi$ is the fraction of NS born with  magnetic fields $B
\geq 10^{14}$ G, which from our simulations is about $8\%$. Thus,
$\lambda_p = 3.9\times 10^{-4}\, \mathrm{M}_{\odot}^{-1}$.

The total spectral gravitational energy emitted by a rotating neutron star is given by 
\begin{equation}
\frac{dE_{gw}}{d\nu} = \frac{dE_{gw}}{dt}\mid\frac{dt}{d\nu}\mid
\end{equation}
where the frequency evolution, according to the analysis of Section 2.2, is controlled by the 
magnetic breaking. In this case, one obtains (RP01b)
\begin {equation} 
\frac{dE_{gw}}{d \nu} = K \nu^3 = \frac{512G}{5 c^5} \pi^6 \varepsilon^2
I_{zz}^2 \frac{\tau_0}{P_0^2} \nu^3.
\end {equation}
Notice that here we have corrected the above equation by the factor of two missing in the original 
equation derived by RP01b. The ratio between the initial period $P_0$ and the magnetic braking
time-scale $\tau_0$  is
 \begin{equation}
\frac{P_0^2} {\tau_0} = \frac{4 \pi^2 R^6}{3I_{zz}c^3} B^2. 
\end{equation}
For magnetars, the ellipticity is given by eq.(1) and in this case the constant $K$ is given by

\begin{equation}
K = \frac{384\pi^4}{5}g^2I^3_{zz}(\frac{R}{G^{1/2}M^2c})^2B^2 \mathrm{sin}^4(\alpha) 
\end{equation}
or, numerically
\begin{equation}
K = 2.0\times 10^{31}g^2I_{45}^3R_{10}^2M_{1.4}^{-4}B_{14}^2 \mathrm{sin}^4(\alpha)
\end{equation}
where $I_{zz}$ is given in units of $10^{45}$ gcm$^2$.

According to our simulations, the average value of the quantity
$<(B_{14}^2\mathrm{sin}^4\alpha)>$ for the galactic population is 32.2, which
will be assumed to be also representative of extra-galactic magnetars.
Notice that if the migration of the magnetic dipole is introduced as
in RP01, the average $<(B_{14}^2\mathrm{sin}^4 \alpha)>$ should be replaced by
$<(B_{14}^2\mathrm{sin}^2\alpha)>$ in eq. 17. However, this
alternative model does not 
change significantly the numerical value of K, since the migration scale time is rather short
($< 10^4$ yr) and, as a consequence, most of the objects will be quickly found in an 
orthogonal configuration.

\section{Results and detectability}

Fig. 5 shows the density parameter $\Omega_{gw}$ as a function of
the observed frequency for two NS models. Model A considers a normal
interior (non-superconducting), having a representative deformation
parameter $g = 13$ (Konno, Obata \& Kojima 2000); model B assumes the existence of a 
type I superconducting interior, characterized by a deformation parameter $g = 520$
(Bonazzola \& Gourgoulhon 1996).

The density parameter $\Omega_{gw}$ reaches a maximum around 1200 Hz and has
a high frequency cut-off at $\sim$ 2500 Hz. The maximum value of the
density parameter is about $6\times 10^{-13}$ in the case of model A and
$\sim 10^{-9}$ if the NS has a superconducting interior.

%
%
\begin{figure}
\resizebox{\hsize}{!}{\includegraphics[width=5in]{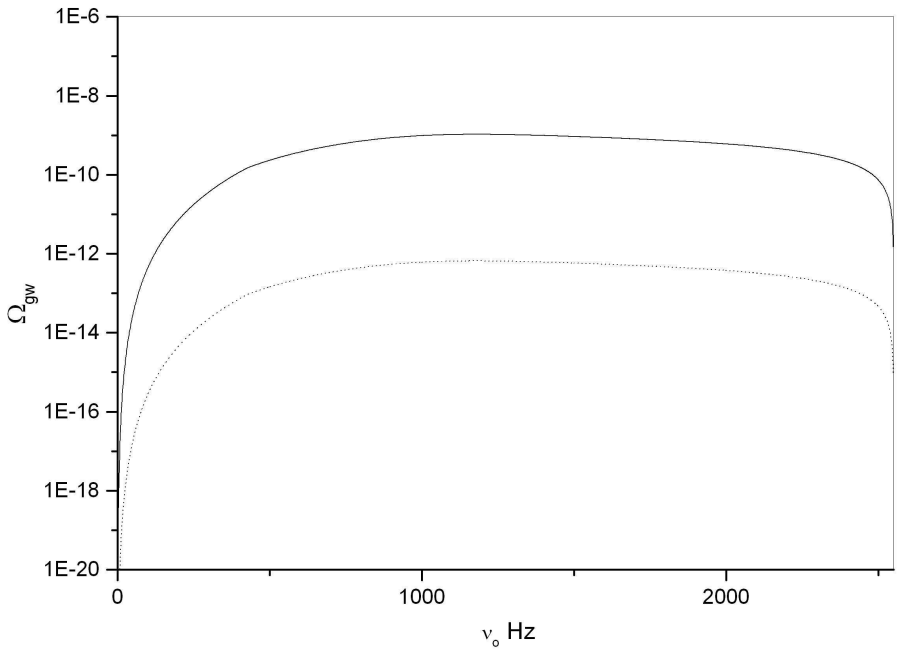}}\hfill
\caption{Density parameter $\Omega_{gw}$ for the stochastic background produced by
magnetars for the two models considered in this work. Model A, solid curve and model B,
dotted curve.}
\end{figure}

If $\bar{\tau}$ is the duration of the signal produced by a single source,
the stochastic background from magnetars is expected to be continuous
since the duty cycle satisfies $D=\int_{0}^{\infty}dR(z)\bar{\tau}(1+z) >> 1$. 
Because it obeys Gaussian statistics and can be confounded with
the instrumental noise background of a single detector, the optimal detection 
strategy is to cross correlate the output of two (or more) detectors, assumed to have independent 
spectral noises. The cross correlation product is given by (Allen \& Romano, 1999):

\begin{equation} 
Y=\int_{-\infty}^\infty \tilde{s_1}^*(f)\tilde{Q}(f)\tilde{s_2}(f) df
\end{equation}
where
\begin{equation}
\tilde{Q}(f)\propto \frac{\Gamma (f) \Omega_{\rm gw}(f)}{f^3P_1(f)P_2(f)}
\end{equation}
is a filter that maximizes the signal to noise ratio ($S/R$). In the above equation, $P_1(f)$ and 
$P_2(f)$ are the power spectral noise densities of the two detectors
and $\Gamma$ is the non-normalized overlap
reduction function, characterizing the loss of sensitivity due to
the separation and the relative orientation of the detectors. 
The optimized $S/N$ ratio for an integration time $T$ is given by
(Allen, 1997):

\begin{equation}
(\frac{S}{N})^2 =\frac{9 H_0^4}{8 \pi^4}T\int_0^\infty
df\frac{\Gamma^2(f)\Omega_{\rm gw}^2(f)}{f^6 P_1(f)P_2(f)}.
\end{equation} 

In the literature, the sensitivity of detector pairs  is usually given in terms of the minimum
detectable amplitude for a flat spectrum ($\Omega_{gw}$ equal to constant) 
(Allen \& Romano, 1999)

\begin{equation}
\Omega_{min}=\frac{4 \pi^2}{3H_0^2\sqrt{T}}({\rm erfc}^{-1}(2 \alpha)-{\rm erfc}^{-1}(2 \gamma)) 
 \lbrack \int_0^\infty df \frac{\Gamma^2 (f)}{f^6 P_1(f) P_2(f)}\rbrack^{-1/2}
 \end{equation} 

The expected minimum detectable amplitudes for the main detector pairs, after one
year integration, are given in Table 1, for
 a detection rate $\alpha=90\%$  and a false alarm rate $\gamma=10\%$.

The power spectral densities used for the present calculation can be
fond in Damour et al. (2001).

%
%

\begin{table}
\caption{Expected $\Omega_{min}$ for the main detector pairs, corresponding to a flat
  background spectrum and an integration time T = 1 year, 
a detection rate $\alpha=90 \%$  and a false alarm rate $\gamma=10\%$.
LHO and LLO stand for LIGO Hanford Observatory and  LIGO Livingston Observatory } 
\begin{flushleft}
\begin{tabular}{lcccc}
\noalign{\smallskip}
\hline
\noalign{\smallskip}
& LHO-LHO & LHO-LLO & LLO-VIRGO & VIRGO-GEO  \\
\noalign{\smallskip}
\hline
\noalign{\smallskip}
initial& $4 \times 10^{-7}$ & $4 \times 10^{-6}$ &  $8 \times 10^{-6}$
&  $8 \times 10^{-6}$  \\
advanced & $6 \times 10^{-9}$  & $1 \times 10^{-9}$  & &  \\
\noalign{\smallskip}
\hline
\end{tabular}
\end{flushleft}
\end{table}
$\Omega_{min}$ is of the order of $10^{-6}-10^{-5}$ for the first generation of interferometers
combined as LIGO/LIGO and LIGO/VIRGO. Their advanced counterparts will permit an increase of
two or even three orders of magnitude in sensitivity ($\Omega_{min} \sim 10^{-9}-10^{-8}$).
The pair formed by the co-located and co-aligned LIGO Hanford detectors, for
which the overlap reduction function is equal to one, 
is potentially one order of magnitude more sensitive than the
Hanford/Livingston pair, if instrumental and environmental noises can be removed.
However, the spectrum of magnetars {\it is not flat} and the maximum occurs out of the optimal frequency 
band of ground based interferometers, which is typically around few
hundreds Hz, as in Fig. 6. Considering the co-located and co-aligned advanced LIGO interferometer pair,
we find a signal-to-noise ratio $S/N \sim 3\times 10^{-6}$ and $\sim 5 \times 10^{-3}$  for 
models A and B respectively.
A possible solution to increase the sensitivity at higher frequencies would be to use 
the interferometers in a narrowband (rather than in a wideband) configuration, as has 
been proposed for GEO 600, with the frequency of maximal sensitivity centred around 1 kHz. 
This could be achieved by choosing a high (rather than low)
reflectivity of the signal in the recycling mirror, which would concentrate
the power in a narrow band rather than distribute it in a
wideband. The resonance frequency of the recycling cavity is tunable to
 the desired frequency by shifting the signal-recycling mirror.

%
%
\begin{figure}
\resizebox{\hsize}{!}{\includegraphics[width=5in]{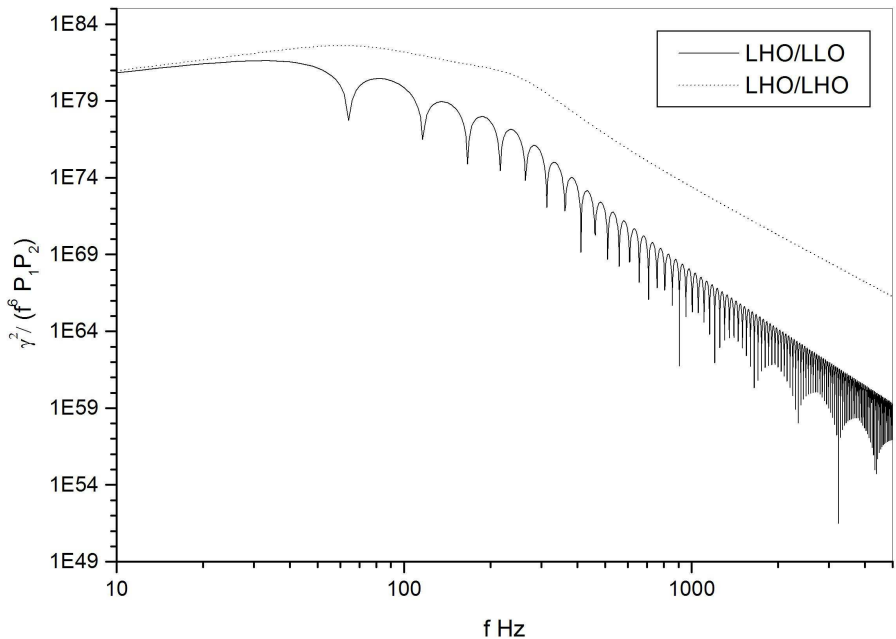}}\hfill
\caption{Integrand of the signal to noise ratio (see eq. 21) for advanced LIGO
  pairs: LHO-LHO (continuous line) and LHO-LLO (dot line)}
\end{figure}

Unless the deformation parameter $g$ is substantially higher than
the present expectations, our results indicate that its contribution to the
gravitational background is not detectable by the
present generation antennas and also probably by their advanced
counterparts. A detection could be possible if the interior of the NS
is constituted by a type II superconductor. Effective quark-quark
interactions in the deconfined core may form a diquark condensate,
which behaves as a superconductor of the second kind Blaschke, Sedrakian
\& Shahabasyan 1999), able to develop quite high fields and to produce a considerable 
magnetic distortion ($g > 10^3$). The deformation of the star can be increased
if the magnetic field has a geometry other than dipolar, since higher order multipoles
will also contribute to the mass quadrupole. This is also true if the toroidal component
is considered since its strength may be a factor of 10 higher than the poloidal
component. Clearly these last possibilities require further work on NS structure
when magnetic fields with complex geometries are considered. 

The background from magnetars may not be detectable as a signal by the
first generations of interferometers but it may behave as noise in the
cosmological background frequency domain.
%
%
\begin{figure}
\resizebox{\hsize}{!}{\includegraphics[width=5in]{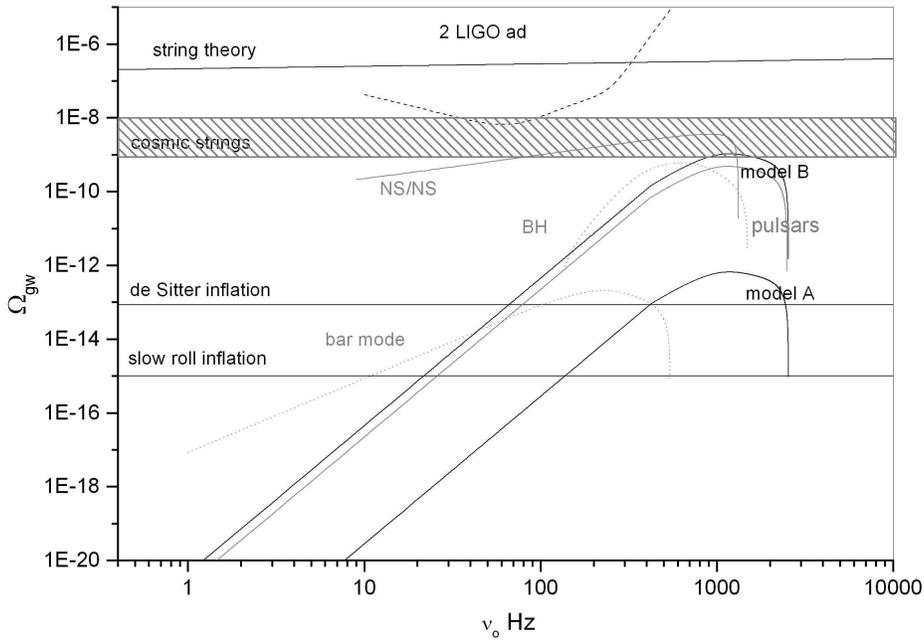}}\hfill
\caption{Theoretical limits for some cosmological gravitational background models and present
estimates for the magnetar contribution. 
Other astrophysical contributions (in grey) are plotted for
comparison: continuous background (continuous line) such as NS/NS coalescence (Regimbau \&
de Freitas Pacheco 2005) or rotating
pulsars (updated from RP01b) and shot
noise (dot line) such as BH ringdown and bar mode due to the
Maclauren/Dedekind transition in young and fast rotating NS (updated from Regimbau 2001).   
The expected sensitivity resulting from the combination
of two advanced interferometers is also shown for comparison.}
\end{figure}

In Figure 7, we have plotted some theoretical predictions for the CGB (Allen 1997; Maggiore 1999;
Buonanno 2004), our estimates for the magnetar contribution along with
other AGBs and the expected sensitivity
of two correlated advanced interferometers. The amplification of vacuum fluctuations at
the transition between the radiation and matter eras, first discussed by Grishchuk (1974 \&
1993) and Starobinsky (1979), is expected to produce a GW background whose spectrum
and amplitude depend strongly on the fluctuation power spectrum developed during the
early inflationary period.

The standard de Sitter inflation predicts a spectrum that
decreases as $~1/f^2$ in the range $3\times 10^{-18}-1\times 10^{-16}$
Hz and then remains constant in a very large band up to frequencies of
the order of MHz.
The Cobe experiment provides an upper bound of $\Omega_{gw} =9\times 10^{-14}$
(Maggiore 1999; Buonanno 2004) for the flat region, which is below the
estimated background from
magnetars in the frequency range $~400-2500$ Hz
for model A and  $~70-2500$ Hz for model B. These
predictions have been questioned recently by Grishchuk (2005).
In the  more realistic scenario of ``slow roll down'' inflation, the
predicted amplitude
is even lower, by almost two orders of magnitude.
A more interesting case arises from the string cosmology since
the expected amplitude is around $\Omega_{gw}=10^{-7}$, well above the expected contribution from
magnetars. However, other choices of the model parameters may lead to substantially
lower amplitudes. Cosmic strings are also able to produce  a CGB in
the frequency range of the magnetar contribution, since the vibrations of these topological defects are sources
of GW (Buonanno 2004).
The spectrum is almost flat in a large frequency band ($10^{-8}-
10^{-10}$) with an 
upper limit for the amplitude of $10^{-9}-10^{-8}$
 according to numerical simulations of a cosmic strings network that satisfy the bound imposed by the timing of
pulsars. As it can be seen from fig. 7, the magnetar contribution from model B is barely
comparable to the predictions of this model at frequencies around 1 kHz. However,
according to a recent study by Damour and Vilenkin (2000, 2001, 2005) the background
from a network of cosmic strings at the end of Brane inflation may be not continuous but
rather a collection of bursts emitted from cusps of oscillating string loops. In this case,
it would be distinct in its statistical properties from the magnetar contribution.

\section{Conclusions}

In the present work, the contribution of magnetars to the stochastic background of
gravitational waves was investigated, under the assumption that these highly
magnetized stars are typical NS born in the high side tail of the initial field
distribution (RP01a). Their geometric deformation is a consequence of such fields.

The statistical properties of the magnetar population such as the magnetic field and
ellipticity distributions and their relative orientation were derived by simulations and then 
by supposing that the extra-galactic population has similar properties. Our new
population synthesis, including new pulsars discovered by the Parkes multibeam survey,
indicates that magnetars (neutron stars born with fields higher than 10$^{14}$ G)
constitute 8\% of the total population.

Two models for the neutron star interior have been considered in our
calculations: in the 
first (model A), the NS has a normal conducting interior, leading to deformation parameters 
$g$ in the range 1-20. Here, the value $g = 13$ was adopted (Konno, Obata \& Kojima 2000). The 
second possibility (model B) considers that the NS develops a type I superconducting 
interior with a typical deformation parameter $g = 520$ (Bonazzola \& Gourgoulhon 1996).

We found that magnetars produce a continuous background (duty cycle $> >$ 1) with a maximum
around 1.2 kHz. This value is slightly dependent on the assumed maximum rotation velocity
of the NS at birth, here taken equal to 1250 Hz (corresponding to a minimum initial rotation period
of 0.8 ms). The different cosmic star formation rates do not
considerably affect the
resulting density parameter $\Omega_{gw}$, whose amplitude depends on the
deformation parameter $g$, e.g., on the conducting properties of the NS interior. Amplitudes 
higher than the maximum found in this work, e.g., $\Omega_{gw} \sim 10^{-9}$, could be expected if 
neutron stars have deconfined cores and a diquark condensate forms a second kind superconducting 
medium. However, estimates of the resulting deformation parameter under these conditions
remain to be done. 
In spite of the maximum amplitude being inside the frequency band of ground based
interferometers, the expected signal is well below the sensitivity of the first generation
of detectors, but exceeds the upper bound derived from COBE data. Comparing our
estimates to the most optimistic expectations of the gravitational background generated
by some possible models, we find that the magnetar contribution does not overwhelm the
relic emission expected from string cosmologies, but could be comparable around 1.0
kHz to the limits estimated for the emission from oscillating cosmic strings. Above $70 -
120$ Hz the magnetar contribution surpass the theoretical limits of the inflationary CGB,
suggesting that searches should remain well below those frequencies.
In spite of the maximum amplitude lying inside the frequency band of ground based interferometers,
the expected signal is well below the sensitivity of the first generation of detectors.

\begin{acknowledgements}
We are grateful to L. Grishchuk and M. Maggiore for discussions during the
preparation of this manuscript.
\end{acknowledgements}



\begin{thebibliography}{}
\bibitem{} Andersson N. and Kokkotas K.D., 2001, Int.J.Mod.Phys. D10, 381
\bibitem{} Allen B., 1997 {\it Relativistic Gravitation and Gravitational
  Radiation} (eds Marck J A and Lasota J P / Cambridge University Press) p. ~373
\bibitem{} Bildsten L., 1998, ApJ 501, L89 
\bibitem{} Bonazzola S. and Gourgoulhon E., 1996, A\&A, 312, 675
\bibitem{} Buonanno A., 2004, gr-qc/0303085
\bibitem{} Chandrasekhar S. and Fermi E., 1953, ApJ 118, 116
\bibitem {} Christensen N., 1997, PRD, 55, 448
\bibitem{} Coccia E. and Fafone V., 1997, Omnidirectional Gravitational Radiation Observatory,
eds. W.F. Velloso, O.D. Aguiar and N.S. Magalh\~aes, World Scientific,
p.113
\bibitem{} Coward D., Burman R.R. and Blair D. 2001, \mnras, 324, 1015
\bibitem{} Damour T., Iyer B.R. and Sathyaprakash B.S., 2001, PRD, 63, 044023
\bibitem{} Damour T. and Vilenkin A., 2000, Phys. Rev. Lett., 85, 3761
\bibitem{} Damour T. and Vilenkin A., 2001, PRD, 64, 064008
\bibitem{} Damour T. and Vilenkin A., 2005, PRD, 71, 063510
\bibitem {} de Bernardis P. et al., 2000, Nature, 404, 995
\bibitem{} de Freitas Pacheco, J.A., 1998, A\&A 336, 397
\bibitem{} de Freitas Pacheco J.A., 2001, in Relativistic Aspects of
  Nuclear Physics, eds. T. Kodama, C. E. Aguiar, H.-T. Elze, F. Grassi, Y. Hama, G. Krein, World Scientific, p. 158
\bibitem{} Duncan R. C. and Thomson C., 1992, ApJ, 392, L29
\bibitem {} Ferrari V., Matarrese S. and Schneider R., 1999, MNRAS, 303, 258
\bibitem{} Ferraro V.C.A., 1954, ApJ 119, 407
\bibitem{} Gal\'tsov D.V. and Tsvetkov V.P., 1984, Phys.Lett. 103A,
  193
\bibitem{} Grishchuk L.P., 1974, Sov. Phys. JETP, 40, 409
\bibitem{} Grishchuk L.P., 1976, Sov. Phys. JETP Lett., 23, 293
\bibitem{} Grishchuk L.P., 1993, CQG, 10, 2449
\bibitem{} Grishchuk L.P., 2005, gr-qc/0504018 
\bibitem{} Grishchuk L.P., Lipunov V.M., Postnov K.A., Prokhorov M.E. and Sathyaprakash B.S., 2001,
Physics-Uspekhi, 44, 1 1993, CQG, 10, 2449
\bibitem {} Hanany S. et al, 2000, ApJ, 545, L5
\bibitem{} Heyl J.S. and Kulkarni S.R., 1998, ApJ 506, L61
\bibitem {} Konno K., Obata T. and Kojima Y., 1999, A\&A, 352, 211  
\bibitem {} Konno K, Obata T. and Kojima Y. 2000, A\&A, 356, 234
\bibitem{} Kouveliotou C., Dieter S., Strohmayer T., van Paradijs J., Fishman G.J., Meegan C.A.,
Hurley K., 1998, Nature 393, 235
\bibitem{} Kouveliotou C. et al., 1999, ApJ 510, L115
\bibitem{} Maggiore M., 2000, gr-qc/0008027
\bibitem{} Maggiore M., 2000, Phys.Report, 331, 283
\bibitem{} Marranghello G.F., Vasconcellos C.Z. and de Freitas Pacheco J.A., 2002, PRD 66, 064020
\bibitem{} Mereghetti S., 1999, in The Neutron Star-Black Hole Connexion, NATO-ASI, p. 13,astro-ph/9911252
\bibitem{} Owen B.J., 2005, astro-ph/0503399
\bibitem {} Palomba C.,  2001, A\&A, 367, 525
\bibitem {} Porciani C. and Madau P., 2001, ApJ, 548, 522 
\bibitem{} Regimbau T. 2001, PhD Thesis, University of Nice
\bibitem {} Regimbau T. and de Freitas Pacheco J. A., 2000, A\&A,
  359, 242 (RP00)
\bibitem {} Regimbau T. and de Freitas Pacheco J. A., 2001a, A\&A,
  374, 182 (RP01a)
\bibitem {} Regimbau T. and de Freitas Pacheco J. A., 2001b, A\&A,
  376, 381 (RP01b)
\bibitem {} Regimbau T. and de Freitas Pacheco J. A., 2005, in preparation
\bibitem {} Schmidt B. et al., 1998, ApJ, 507, 46
\bibitem{} Spergel et al., 2003, ApJS, 148, 175
\bibitem {} Thorne K. S., 1980, Rev.Mod.Phys., 52, 299
\bibitem{} Ushomirsky G., Cutler C. and Bildstel L., 2000, MNRAS 319, 902
\bibitem {} Young M. D., Manchester R. N. and Johnson S., 1999,
  Nature, 400, 848 
\bibitem{} Zimmermann M. and Szedenits E., 1979, PRD 20, 351
\end{thebibliography}
\end{document}